\documentclass[sigconf,nonacm=true]{acmart}

\usepackage{enumitem}
\usepackage{comment}
\usepackage{tikz}
\usetikzlibrary{positioning} 
\usepackage[euler]{textgreek}
\usepackage[normalem]{ulem}
\usepackage{alltt}

\setcopyright{none}
\AtBeginDocument{%
  \providecommand\BibTeX{{%
    \normalfont B\kern-0.5em{\scshape i\kern-0.25em b}\kern-0.8em\TeX}}}

\begin{document}

\title{A GPU Register File using Static Data Compression}


\author{Alexandra Angerd, Erik Sintorn, and Per Stenstr{\"o}m}
\affiliation{%
  \institution{Department of Computer Science and Engineering\\Chalmers University of Technology}
  \city{G{\"o}teborg}
  \country{Sweden}}
\email{{angerd,erik.sintorn,per.stenstrom}@chalmers.se}

\renewcommand{\shortauthors}{Angerd et al.}

\begin{abstract}
GPUs rely on large register files to unlock thread-level parallelism for high throughput. Unfortunately, large register files are power hungry, making it important to seek for new approaches to improve their utilization.

This paper introduces a new register file organization for efficient register-packing of narrow integer and floating-point operands designed to leverage on advances in static analysis. We show that the hardware/software co-designed register file organization  yields a performance improvement of up to 79\%, and 18.6\%, on average, at a modest output-quality degradation.
\end{abstract}




\maketitle

\section{Introduction}
Modern GPUs provide a high throughput by enabling massive thread-level parallelism (TLP). Large register files are needed to provide fast context-switching between threads, and GPUs rely on ever larger register files in order to further increase thread-level parallelism (TLP)~\cite{7448930,8029511}. The Fermi architecture, a few generations back, had a register file size of about 131 KB per streaming multiprocessor (SM), consuming 13.4\% of the total dynamic power~\cite{Leng:2013:GEE:2485922.2485964}. With 15 SMs, it sums up to about 2 MB register storage in total. In contemporary architectures, such as NVIDIA's Turing, the total register file size sums up to approximately 18 MB~\cite{turing-whitepaper}; a ninefold increase!

 TLP can be increased by leveraging a more efficient utilization of the register resources to decrease the per-thread register footprint. In a conventional register file, the operands are stored at a granularity of 32 bits. However, it has been shown that many operands in GPU workloads require significantly less space. For example, a large portion of the integer operands is narrow~\cite{8029511,6522330}. Furthermore, in the case of floating-point operands, their precision can be significantly reduced with a negligible impact on quality~\cite{Sathish:2012:LLM:2370816.2370864}, especially if the operands are allowed to have different precisions~\cite{Angerd:2017:FAC:3154814.3151032}.

Previous work~\cite{8029511,6522330,AsghariEsfeden:2019:CCO:3297858.3304026} propose register-file organizations which pack operands by establishing the number of significant bits, i.e. the \textit{bitwidth}, at run time using zero/one-detection logic to remove redundant sign extension bits. However, this approach does not work for floating-point data. Still, it has been shown that the precision of many floating-point operands can be reduced substantially~\cite{Angerd:2017:FAC:3154814.3151032} with negligible impact on the quality of the application output. In addition, the precision of a floating-point operand cannot be decided at \textit{run time}, since it is not possible to know what impact reduction of precision of a specific operand has on the end result. 

In this paper, we propose, for the first time, a register file organization capable of storing operands at a fine granularity \textit{regardless} of the data type. Our evaluation shows that in order to achieve significant reduction of the register footprint, \textit{both} integer and floating-point data have to be considered. To the best of our knowledge, our approach is the first to support mixed packing of both integers and floating-point numbers. We do this by annotating the bitwidth needed by each operand at the instruction level, at compile time. To detect narrow integers, we leverage static range analysis~\cite{QuintaoPereira:2013:FLT:2495258.2495937}. To determine the precision of each float, we leverage a method that tunes the precision to meet a user-defined output quality threshold~\cite{Angerd:2017:FAC:3154814.3151032}. The annotations are taken into consideration to achieve a dense register allocation. At run time, our proposed register file uses a configurable indirection table to store the location of each operand.

Our approach is inspired by the idea of a configurable indirection table given by Angerd et al.~\cite{Angerd:2017:FAC:3154814.3151032}. They assume the existence of a register file with support for low-precision floats. However, they do not consider any support for narrow integers. Furthermore, they do not present any microarchitectural design of such a register file, nor how it could be integrated into a GPU pipeline, not even for floats. Hence, how to design a compiler-assisted register file which supports both integer and floating-point data remains unsolved. In particular, a microarchitectural design of such a register file organization implies a number of challenges, which we address in this paper: First, the indirection table needs to be consulted for each and every register access. Since this access is on the critical path, it introduces latency which can have an adverse impact on performance. Second, the indirection table must be able to handle multiple accesses per cycle, since several register accesses have to be carried out simultaneously. Third, conversions between different floating-point formats are necessary. We mitigate these challenges by presenting an indirection table microarchitecture which matches the throughput of the register file, as well as conversion units capable of carrying out floating-point format changes in one cycle. Our evaluation shows that our approach yields a performance improvement of up to 79\% and 18.6\%, on average, compared to a register file which only supports an operand granularity of 32 bits.

Our contributions in this paper are the following:
\begin{itemize}[leftmargin=*,noitemsep]

\item A new register file organization for GPUs which, unlike previously proposed solutions, is capable of supporting narrow operands \textit{regardless} of data type (integer and floating point data).
\item A new concept for efficient packing of narrow operands, which is built upon static bitwidth analysis co-designed with a new register file concept.
\item An evaluation of the proposed microarchitectural design, which shows a performance improvement of up to 79\%, and 18.6\% on average, when allowing for a slight output quality loss.
\end{itemize}

The rest of the paper is organized as follows. Section~\ref{motivation} provides a motivational example. Section~\ref{regfile} introduces the baseline GPU architecture and our proposed register file organization. Section~\ref{framework} presents the static approach we use to reduce operand bitwidths. Section~\ref{method} describes the methodology used to derive the results in Section~\ref{results}. We discuss the implications on other architectures in Section~\ref{discussion}. Finally, we put our work in the context of related work in Section~\ref{relatedwork} before we conclude in Section~\ref{conclusion}.

\section{Motivation}
\label{motivation}
In this section, we show the performance improvement obtained by increasing TLP through improved register file utilization for a kernel called IMGVF, included in the Leukocyte application from the Rodinia benchmark suite~\cite{5306797}. The core idea is to reduce the \textit{register pressure}, that is,  the maximum number of fixed-size registers needed by a thread, by reducing the bitwidth of the operands. This way, the register file can accommodate more threads. For IMGVF, the original register pressure is 52 registers.

The kernel's register pressure directly affects how many threads are allocated to each GPU core. In the CUDA programming model, warps consist of 32 threads which are further bundled into blocks, whose size are kernel-specific and decided by the programmer. The assignment of threads to a core is done at the granularity of blocks. A Fermi GPU has 32,768 registers per core, and can support up to 48 warps to be active, simultaneously. However, the block size for IMGVF is ten warps with a register pressure of (52 x 32 x 10 =) 16,640 registers. Hence, only one block fits in the register file; the register usage severely limits the achievable TLP. We refer to this limit as occupancy, i.e. the ratio of active warps to the maximum number of warps supported by the core. 

\begin{table}
    \centering
    \caption{Register pressure, occupancy, and IPC of the baseline, when using either one or both parts of the framework, and when artificially increasing the occupancy.}
    \label{table_motivation}
\resizebox{\linewidth}{!}{
    \begin{tabular}{|c|c|c|c|}
        \hline
          & \textbf{Register Pressure} & \textbf{Occupancy}  & \textbf{IPC}\\ \hline
         \textbf{Original}                      & 52 & 21\%     & 196\\ \hline
         \textbf{Narrow integers}               & 46 & -        & -\\ \hline
         \textbf{Narrow floats}                 & 36 & -        & -\\ \hline
         \textbf{Narrow integers + floats}      & 29 & 62.5\%   & 352\\ \hline
         \textbf{Artificial occupancy increase} & 52 & 62.5\%   & 377\\ \hline
    \end{tabular}
}
\end{table}

To reduce the register pressure and increase the occupancy, we run the application through the static analysis framework (described in detail in Section~\ref{framework}). It consists of two parts: A static analysis framework  1)  that based on ~\cite{QuintaoPereira:2013:FLT:2495258.2495937} finds the required number of bits for each integer operand and 2)  that based on ~\cite{Angerd:2017:FAC:3154814.3151032} establishes the precision for each floating-point operand given a user-specified quality metric and threshold. Here, we have specified the quality metric and threshold such that no deviation from the original output is allowed. Table~\ref{table_motivation} reports the original register pressure, when each framework is applied in isolation, and when both frameworks are used. When both frameworks are used, the register pressure is lowered from 52 to 29 registers. As a result, three blocks (30 warps) can now fit into the register file simultaneously, as opposed to one block in the original case. Hence, the occupancy is increased from 21\% to 62.5\%.

To confirm that an increase in occupancy would unlock more TLP and higher performance, we run the application with a simulated Fermi GPU, using GPGPU-Sim~\cite{4919648} (details are provided in Section~\ref{method}), both using the original occupancy and the occupancy reached with our compression technique by increasing the size of the simulated register file. The result is presented in Table~\ref{table_motivation}; with a higher occupancy, the Instructions Per Clock (IPC) is increased by 91\%. 

We also modify GPGPU-Sim to include the microarchitectural structures needed to implement the proposed register file using an indirection table approach described in Section~\ref{regfile}. As seen in Table~\ref{table_motivation}, the increase in IPC is close to what can be achieved by artificially allowing for higher occupancy.

\section{New Register File Organization}
\label{regfile}
In this section, we first describe the baseline GPU in Section~\ref{gpu_background}. Then, in Section~\ref{indir_implementation}, we describe the microarchitectural implementation of the new register file organization.

\subsection{GPU Baseline Architecture}
\label{gpu_background}

Our baseline microarchitecture resembles NVIDIA's Fermi architecture~\cite{fermiwhitepaper} used in several recent studies~\cite{Lee:2015:WEP:2749469.2750417,Leng:2013:GEE:2485922.2485964}. The threads in each warp are executed in lockstep, but with different register values in a SIMD-fashion. Hence, threads are scheduled to execution units at warp granularity. 

The GPU has 15 cores called Streaming Multiprocessors (SMs), which share an L2 cache. Each SM has a private L1 cache, a local shared memory, and a dedicated texture cache. Each SM also has two warp schedulers, which together can schedule two instructions from different warps simultaneously. One SM also comprises two Single Precision Units (SPUs), one Special Function Unit (SFU), and one memory (LD/ST) unit. All units execute at the granularity of one \textit{warp instruction} (32 lock-stepped \textit{thread instructions}). The SPUs execute all types of instructions except for built-in trigonometric and logarithmic operations, which are executed using the SFUs. The LD/ST unit carries out memory operations.  

\begin{figure}
\centering
\includegraphics[height=\linewidth,angle=-90]{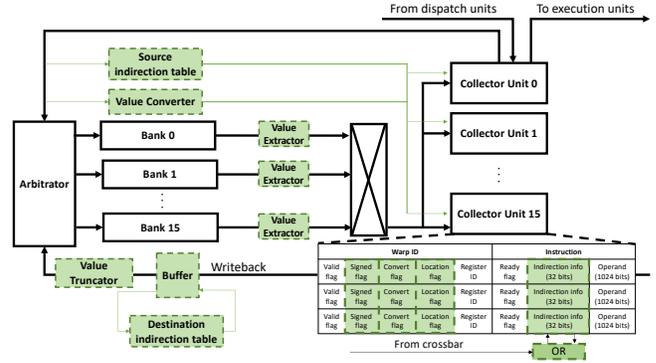} 
\caption{Baseline operand collector design with proposed extensions (in green).}
\label{newopndcoll}
\end{figure}

To hide idle cycles caused by hazards and memory access latency, the SM keeps a large number of warps in flight supporting fast context-switches by a large register file. To provide a large bandwidth and the appearance of being multi-ported, the baseline register file uses an \textit{operand collector} (see Figure~\ref{newopndcoll}). The register file is split into 16 banks, each with 64 entries, 1024 bits wide, with one read port and one write port per bank. Because each warp is executed in lockstep, the registers are stored in vectors of 32 \textit{thread registers}, forming a \textit{warp register}. A register file access applies to a  warp register. To maximize throughput, an arbitrator is associated with the operand collector to distribute the requests from all collector units (CUs) to maximize register-bank accesses in each cycle. 


A warp instruction is allocated to one of the CUs in the operand collector. The \textit{valid} flags in the CU are set indicating which operands to fetch from the register file. The operands are then queued at the arbitrator. Since the arbitrator is optimized for throughput, and not for individual warp latency, only one operand can be collected to each CU in each cycle. Hence, it may take a few cycles before all operands for a certain warp instruction are collected. When all operands for a warp instruction are collected, i.e., when all \textit{ready} flags in one CU are set, the warp instruction is ready to be forwarded to the execution units.

\subsection{Proposed Register File Organization}
\label{indir_implementation}
\begin{figure}
 \centering
\includegraphics[height=\linewidth,angle=-90]{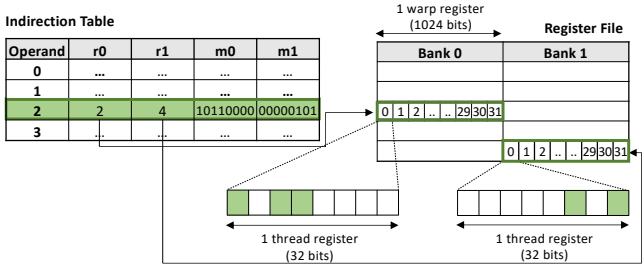} 
\caption{Each thread register is divided into eight slices. These are accessed through an indirection table.}
\label{slices}
\end{figure}

To store operands at a fine granularity, each thread register is divided into \textit{slices} (4 bits each to efficiently support the floating-point format described in Section~\ref{floatingpoint}), with an operand comprising data contained in multiple slices (see Figure~\ref{slices}). An \textit{indirection table} points out in which registers and which slices the operand is stored. The allocation of slices to each operand is static for each kernel (to be described in Section~\ref{framework}), so the configuration of the indirection table is different for each kernel. Changes to the baseline register file are confined to the operand collector as the indirection table keeps all information needed to access individual registers. 

\subsubsection{Overview of Operand Collector}
The changes to the operand collector comprises the green blocks in Figure~\ref{newopndcoll}.

When a register read instruction is allocated to a CU, it collects the location information for each operand from the \textit{Source indirection table}. The operand is then queued to the arbitrator as before. The data returned from the register file is compressed, i.e., only some slices of the returned data contain valid data. The \textit{Value Extractor} rearranges the slices such that the data is properly aligned, and the aligned data is returned to the operand collector. It also sign-extends the operand if it is an integer. If the operand is a narrow float, however, it needs to be extended to single precision by the \textit{Value Converter} before being forwarded to the execution units.

Register write instructions collect location information from  the \textit{Destination indirection table}. However, bank conflicts might occur in the indirection table so a buffer temporarily storing conflicting operands is added. As the probability of a conflict is small (the writeback bus is three operands wide, and the number of banks in the indirection table is 16) the number of buffer entries is negligible; one entry corresponds to one warp-register, which is on the order of 0.1\% of the size of the register file. To store a float operand which takes up less than 32 bits, it is converted to lower precision. Then, the operand is aligned to its corresponding placement inside the physical register using a \textit{Value Truncator}.

\subsubsection{The Indirection Tables}
Since the source indirection table is on the critical path, it is vital that it has the same throughput as the register file. To guarantee this, the organization of the indirection table is similar to that of the operand collector: the SRAM cells are divided into 16 banks. A separate arbitrator distributes read requests in such a way that as many banks as possible are accessed each cycle. We assume 256 architectural registers, where the indirection table has to store 32 bits for each of them. A detailed area analysis of the indirection table and the other structures in the proposed register file is provided in Section~\ref{overhead_analysis}. 

To avoid contention we introduce separate, yet identical, indirection tables for the read and the write paths.

\subsubsection{Value Extractor}
\begin{figure}
\centering
\includegraphics[height=\columnwidth,angle=90]{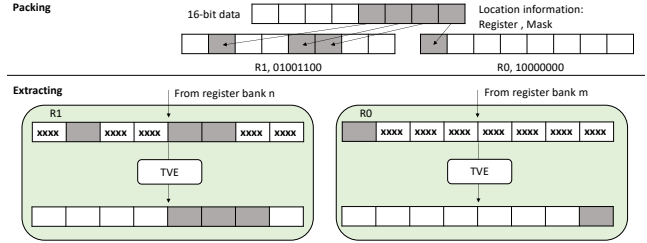}
\caption{Example: a 16-bit float is split into two separate registers. After a fetch, the TVEs extract and align the data.}
\label{packingextracting}
\end{figure}

When the content of a physical register is read from the register file, it is expected that only a few slices contain the required operand. These slices need to be extracted from the rest of the data. As shown in the example scenario of Figure~\ref{packingextracting}, the data of a 16-bit float operand is placed within two separate physical registers: data slice 0 is placed in slice 7 in physical register r0, while data slices 1, 2, and 3 are placed in slices 2, 3, and 6 in physical register r1. To restore the data, the thread value extractor (TVE) rearranges the slices, and sets unused slices to zero. Later, when both physical registers are fetched, the two parts are merged into a complete operand using an OR operation. 

\begin{figure}
\centering
\includegraphics[height=\columnwidth,angle=90]{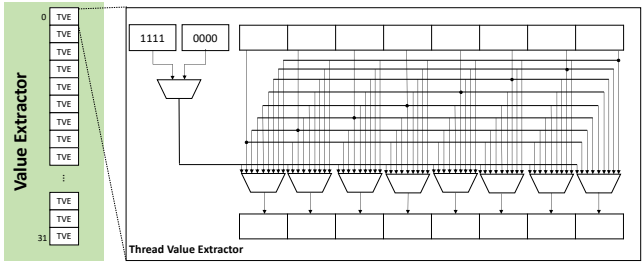}
\caption{The value extractor includes 32 TVEs. Each TVE includes eight 9:1-multiplexers, which select among the input slices and a nibble of zeros or ones.}
\label{tve}
\end{figure}

As Figure~\ref{tve} shows, each value extractor consists of 32 parallel Thread Value Extractors (TVEs). Each of them carries out the extraction for one thread register and it consists of eight 9-to-1 and one 2-to-1 multiplexer. The mask together with some logic gates, connected to the multiplexer select lines, decides the placement of the input slices, zeros, and ones in the output. The 2-to-1 multiplexer decides whether the value should be padded with zeros or ones, depending on the type of operand. A float or unsigned integer should always be padded with zeros, while a signed integer is simply sign-extended.

\subsubsection{Extended Collector Unit}
The CU is extended with four fields per operand, as shown in the lower-right part of Figure~\ref{newopndcoll}. The first is a bit which indicates whether the operand is signed or not. The second is a \textit{convert info} flag, which indicates whether the operand is a float which needs to be converted. The third is a \textit{location} flag, which indicates whether the operand location has been fetched. The fourth is an \textit{indirection info} field, which is filled by an access to the indirection table.

The CU is also extended with a 1024-bit OR-gate, which is used if the operand is split into two different physical registers. The first part that is fetched is simply placed into the operand field. When the second part arrives, it is OR'ed with the data in the operand field to form a complete operand. 


\subsubsection{Value Converter}
\label{value_converter}
The Value Converter (VC) extends low-precision float operands to single-precision floats. Since two instructions can be scheduled in each cycle, and each instruction has up to three source operands, up to six conversions need to be carried out in each cycle to maintain the maximum throughput. Hence, the VC consists of six parallel Warp Value Converters. Each of these, in turn, consists of 32 parallel Thread Value Converters.

The low-precision format we use mimics the IEEE 754 standard, with support for plus/minus infinity and not-a-number values. During format conversion, denormals are truncated to zero, which is safe as the same simplification is made in the precision selection step described in Section~\ref{floattuning}.

\begin{figure}[h]
\centering
\includegraphics[height=\columnwidth,angle=90]{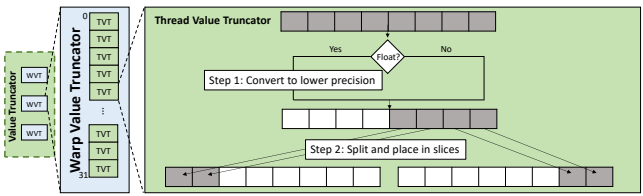}
\caption{The value truncator converts the operand to lower precision and places the data into its assigned slices.}
\label{vt}
\end{figure}

\subsubsection{Value Truncator}
Before an operand with narrow bitwidth is written back into the register file, it has to be adjusted to its assigned slices. This is carried out by the Value Truncator depicted in Figure~\ref{vt}, which comprises three Warp Value Truncators (WVTs). This is because we assume that the writeback bus is three instructions wide, as modelled by GPGPU-Sim~\cite{4919648}. Similar to the WTC, each WVT consists of 32 smaller units called Thread Value Truncators (TVTs). Each TVT carries out the required steps before the operand can be written back. In Step 1, if the operand is a narrow float, it is converted to lower precision. If not, this step is skipped. In Step 2, the data is placed within its corresponding register slices. This procedure is the same as in Figure~\ref{tve}, but with another set of logic for the select lines. Since an operand can be split and placed into two physical locations, two thread value extractors are needed.

In the last step, VTs forward compressed data together with the masks to the register file. At writeback, only the bit lines corresponding to the mask are activated, so as to not overwrite the data in the other slices.

\subsubsection{Pipelining}
\begin{figure}
\centering
\includegraphics[height=\columnwidth,angle=90]{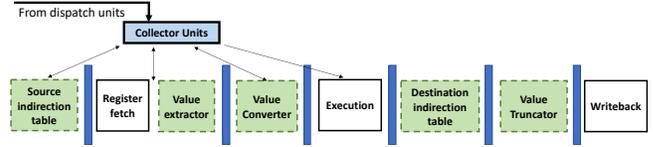}
\caption{Proposed extension of the pipeline (in green).}
\label{pipeline}
\end{figure}

To maintain the baseline clock speed, we modify the pipeline according to Figure~\ref{pipeline}, where the stages marked in green are added to that of the baseline marked in white. In the unmodified pipeline, the operand collector is in charge of sending all its operands to the register fetch stage, and not passing the instruction to the execution stage before all of its operands are collected. In our modified pipeline, the operand collector is also responsible for synchronizing the accesses to the source indirection table, and sending floats to the value converter. We assume all new stages can be carried out in one clock cycle, as will be justified in the next section.

\subsubsection{Timing}
\label{timing}
The indirection table has the same organization as the register file, so we assume the same timing with a maximum throughput of 16 accesses per cycle. 

We estimate the propagation delay of the value converter using Catapult C together with the NanGate 45 nm Open Cell Library by synthesizing it to the register transfer level (Note: Fermi is implemented in 40 nm). A critical path analysis shows that the delay is well within a Fermi clock cycle (0.71 ns). Since the converter has six parallel units, we assume a throughput of six conversions per cycle.

The value extractor has a shallow critical path of one multiplexer. Therefore, we assume it can be carried out within a register-read cycle, and no additional cycles are added.

At writeback, destination operands are looked up in the destination indirection table and, if necessary, truncated using the value truncator. The destination indirection table is identical to the source indirection table, and consequently we assume the same timing. Furthermore, we assume that the value truncator has a similar propagation delay as the value converter. Hence, the minimum writeback delay is two cycles if conversion is needed, and one cycle otherwise. However, in Fermi, the writeback bus is three operands wide, which means that bank conflicts are possible in the destination indirection table. To account for these, we pessimistically model the additional propagation delay as three cycles for \textit{all} operands.


\section{Static Analysis Framework} 
\label{framework}
\begin{figure}
\centering
\includegraphics[height=0.9\columnwidth,angle=90]{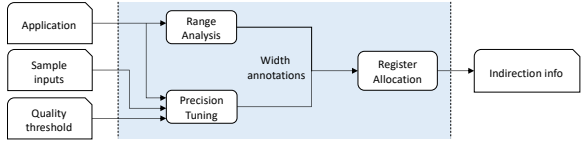}
\caption{Overview of the static framework.}
\label{overview_framework}
\end{figure}

The static analysis framework (see Figure~\ref{overview_framework}) comprises three steps: a range analysis step (Section~\ref{integertuning}), which identifies and reduces the bitwidth of integer operands, a precision-reduction step (Section~\ref{floattuning}), which tunes the precision of the floating-point operands, and a register allocation step (Section~\ref{regalloc}). The range analysis and the precision-reduction steps find and annotate all operands with their needed bitwidths. The register allocation step assigns a suitable number of slices to each operand. Before execution of a kernel, the kernel-specific indirection information is loaded into the indirection table.

\subsection{Floating-Point Precision Tuning}
\label{floattuning}
To reduce the bitwidth of floating-point operands, we employ a method proposed by Angerd et al.~\cite{Angerd:2017:FAC:3154814.3151032}. Their precision-tuning method is a heuristic whose goal is to identify how much the precision of each floating-point value can be reduced while meeting a specified quality threshold. To achieve this, it uses as input an application, a quality threshold, and a number of application sample inputs. These inputs are used to determine how much the precision of each floating-point value can be reduced to meet the quality requirement. It then recursively explores how much the precision of each floating-point value can be reduced. This is carried out at the instruction level, where the instructions are in Single Static Assignment (SSA) form, meaning that each value corresponds to one single value definition. Each SSA register is then annotated with a bitwidth which meets the targeted quality threshold. Obviously, since this approach is data driven, it relies on a domain expert to provide a set of representative sample inputs. No quality guarantees are given for inputs outside of the set.

\subsection{Static Range Analysis}
\label{integertuning}
\begin{figure}
  \begin{tabular}{c  c}
\begin{minipage}[b]{0.32\linewidth}
\begin{alltt}
\scriptsize
k = 0
\textbf{while} k < 50\{
  i = 0
  j = k
  \textbf{while} i < j\{
    \textbf{print} k
    i = i + 1
    k = k + 1
  \}
\}
\textbf{print} k
\end{alltt}
\end{minipage}
&
\begin{minipage}[b]{0.6\linewidth}
\begin{tikzpicture}[thick,scale=0.75, every node/.style={transform shape}]
]
\small
\node   (BB1)       [text width=3cm,align=center] 
{\texttt{k\textsubscript{1} = \straightphi (k\textsubscript{0}, k\textsubscript{2}) \\ k\textsubscript{1} < 50?}};
\node   (BB0)  [left=of BB1,xshift=0.75cm]     
{\texttt{k\textsubscript{0} = 0}};
\node   (BB2)       [text width=3cm,align=center,below= of BB1,yshift=0.6cm] 
{\texttt{ k\textsubscript{t} = k\textsubscript{1}$\cap$[$-\infty$,49]\\
i\textsubscript{0} = 0\\
j\textsubscript{0} = k\textsubscript{t}
}};
\node   (BB3)       [below= of BB0] 
{\texttt{print k\textsubscript{f}
}};
\node   (BB4)       [text width=3cm,align=center,below= of BB2,,yshift=0.6cm] 
{\texttt{i\textsubscript{1} = \straightphi(i\textsubscript{0},i\textsubscript{2})\\
i\textsubscript{1} < j\textsubscript{0}?
}};
\node   (BB6)       [below right= of BB4,,xshift=-1.9cm,yshift=0.5cm] 
{\texttt{k\textsubscript{2} = k\textsubscript{t} + 1
}};
 \node   (BB7)       [text width=3cm,align=center,below left= of BB4,xshift=2cm,yshift=0.6cm] 
{\texttt{print k\textsubscript{t} 
\\i\textsubscript{2} = i\textsubscript{1} + 1
}};

\draw[->] (BB0) -- (BB1);
\draw[->] (BB1) -- node[near start,right] {t} (BB2);
\draw[->] (BB1) -- node[near start,left,xshift=-0.2cm] {f} (BB3);
\draw[->] (BB2) -- (BB4);
\draw[->] (BB4) -- node[near start,left,xshift=-0.2cm] {t}(BB7);
\draw[->] (BB4) -- node[near start,right,xshift=0.2cm] {f}(BB6);

\draw[->] (BB7) |- (BB4);
\draw[->] (BB6) |- (BB1);
\end{tikzpicture}
\end{minipage}

\\
    \textbf{ (a) } & \textbf{ (b) }\\
\begin{minipage}[b]{0.32\linewidth}
\begin{alltt}
\scriptsize
I[k\textsubscript{0}] = [0,0]
I[k\textsubscript{1}] = [0,50] 
I[k\textsubscript{2}] = [1,50] 
I[k\textsubscript{t}] = [0,49] 
I[k\textsubscript{f}] = [50,50] 
I[i\textsubscript{0}] = [0,0] 
I[i\textsubscript{1}] = [0,49] 
I[i\textsubscript{2}] = [1,50] 
I[j\textsubscript{0}] = [0,49] 
\end{alltt}
\end{minipage}
    & 
    
\begin{minipage}[b]{0.55\linewidth}
\begin{alltt}
\scriptsize
I[k] = \(\bigcup\) I[k\textsubscript{x}] = [0,50] 
I[i] = \(\bigcup\) I[i\textsubscript{x}] = [0,50]
I[j] = \(\bigcup\) I[j\textsubscript{x}] = [0,49]

k : 6 bits
i : 6 bits
j : 6 bits
\end{alltt}
\end{minipage}
    
    \\
    \textbf{ (c) } & \textbf{ (d) }\\
  \end{tabular}
\caption{The steps of the static range analysis. (a): Example program. (b): CFG in e-SSA-form. (c): Ranges in e-SSA-form. (d): Range and required bitwidth of each original variable.}
  \label{static}
\end{figure}
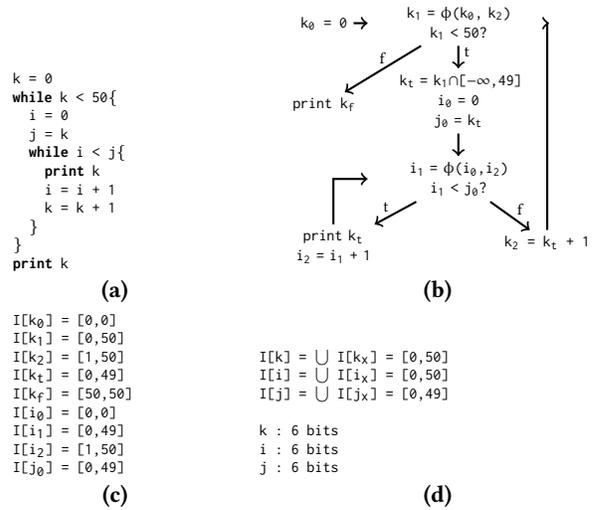

To detect narrow integers offline, we propose to use static range analysis~\cite{QuintaoPereira:2013:FLT:2495258.2495937}. Originally, range analysis was used to secure programs against integer overflows. However, we use static range analysis to determine the number of bits needed for each integer operand. This is carried out at the instruction level. The steps taken in the analysis are shown in Figure~\ref{static}: First, the program is converted into a control flow graph (CFG) which uses a representation called Extended SSA (e-SSA) form. This makes it possible to capture inequalities enforced by control flow dependencies. E.g., the code in Figure~\ref{static}a is converted into the CFG in Figure~\ref{static}b, where the first branch produces two versions of variable \texttt{k}: \texttt{k\textsubscript{t}} which is below 50, and \texttt{k\textsubscript{f}} which is greater than or equal to 50.
Next, the CFG is fed into the range algorithm~\cite{QuintaoPereira:2013:FLT:2495258.2495937}. It creates constraints based on the CFG, analyzes them, and outputs a range for each e-SSA register (Figure~\ref{static}c). Finally, we merge the ranges of all e-SSA registers which belongs to each original variable by finding the union of their ranges, as shown in Figure~\ref{static}d. Finally, we determine how many bits are needed to describe this range. 

\subsection{Register Allocation}
\label{regalloc}
To allocate registers, we make use of an existing algorithm~\cite{Angerd:2017:FAC:3154814.3151032}. We extend the algorithm to consider the width annotations from the range analysis as well as the precision reduction step, and assign a sufficient amount of slices to each operand. The output contains information about the location within the register file for each operand (denoted "indirection info" in Figure~\ref{overview_framework}): a register name points out which physical register to access, and an 8-bit mask shows which slices within the physical register are allocated for the architectural register. In addition, to minimize fragmentation, each architectural register can be split into two parts and placed into the slices of two \textit{different} physical registers, which is why each entry in the indirection table in Figure~\ref{overview_framework} has two physical registers, $r0$ and $r1$, and two masks, $m0$ and $m1$.

\section{Evaluation Methodology}
\label{method}
\subsection{Simulation Set-Up}
We evaluate the impact of our proposed design by modifying GPGPU-Sim~\cite{4919648}. Table~\ref{gpgpusim-config} summarizes the settings, which correspond to the configuration of a Fermi GTX 480 GPU. While relatively old, this baseline is widely used in GPU microarchitecture research (e.g. ~\cite{regmutex, Koo:2017:APC:3079856.3080239, AsghariEsfeden:2019:CCO:3297858.3304026}). The reason for that is that the basic SM pipeline in contemporary GPUs is similar to the one in Fermi. Therefore, our proposal is also applicable to newer architectures: the thread-to-register ratio has not changed much. Register shortage remains a problem. In Section~\ref{discussion} we give further insight into how our proposal scales to other architectures than Fermi.

GPGPU-Sim simulates NVIDIA's instruction set PTX. We use the framework described in Section~\ref{framework} to annotate each PTX register with a bitwidth. Then, the register allocator outputs the indirection table contents in the form of register IDs and masks. This information is then uploaded to GPGPU-Sim and consulted before any register access is carried out. 

\begin{table}
    \caption{Summary of GPU parameters.}
    \label{gpgpusim-config}
\resizebox{\linewidth}{!}{
    \begin{tabular}{|c|c||c|c|}
        \hline
        \textbf{Parameter}  &\textbf{Value} &  \textbf{Parameter}&\textbf{Value} \\
         \textbf{(per GPU)}     & & 
         \textbf{(per SM)}     & \\ \hline 
         Clock Frequency        & 1400 MHz 
         & Warp Schedulers        & 2 \\ \hline
         SMs                    & 15     
         & Max Warps              & 48 \\ \hline
         
         Scheduling Policy      & Greedy then oldest 
         & Thread Registers       & 32768\\ \hline
         L2 cache               & 786 KB 
        &Register Banks         & 16 \\ \hline
         &&Register Bank Width    & 1024 bits \\ \hline
         &&Entries / Bank         & 64 \\ \hline
         &&Operand Collectors     & 16 \\ \hline
         &&L1 cache               & 16 KB\\ \hline
         &&Shared memory          & 48 KB\\ \hline
    \end{tabular}
    }
\end{table}

The PTX instruction set is an intermediate representation compiled by ptxas, the proprietary NVIDIA backend compiler, into the target assembly code. Since we carry out annotations and register allocation directly on PTX, our register usage deviates from what ptxas reports. In all cases, our liveness analysis reports slightly more registers than ptxas does, since the PTX assembly code is not fully optimized. Hence, our register usage is an overestimation compared to what is required in the executed assembly code.

\subsection{Floating-Point Formats}
\label{floatingpoint}
The IEEE 754 standard defines five floating-point precision formats, of which three are supported by modern GPUs: double, single, and half precision (64, 32, and 16 bits respectively)~\cite{turing-whitepaper}. The formats we consider are listed in Table~\ref{floatformat}. Besides the standard 32- and 16-bit precision formats, the rest are chosen to approximately maintain the single-precision ratio between the exponent and mantissa bits. We choose this format because prior research~\cite{Angerd:2017:FAC:3154814.3151032} has shown that it outperforms both using only the formats supported by the IEEE 754 standard as well as mantissa truncation in how efficiently each bit is used. 

As none of our benchmarks use double precision, we do not consider precision formats larger than 32 bits.

\begin{table}
    \centering
    \caption{Distribution of bits for each considered floating-point format. All configurations also include a sign bit.}
    \label{floatformat}
\resizebox{0.75\linewidth}{!}{
    \begin{tabular}{|c|c|c|c|c|c|c|c|}
        \hline
         \textbf{Bits, Total} & \textbf{32} & \textbf{28}  & \textbf{24} & \textbf{20} & \textbf{16} & \textbf{12} & \textbf{8} \\ \hline
         \textbf{Exponent bits} & 8 & 7 & 6 & 5 & 5 & 4 & 3 \\ \hline
         \textbf{Mantissa bits} & 23 & 20 & 17 & 14 & 10 & 7 & 4 \\\hline 
    \end{tabular}
}
\end{table}

\subsection{Benchmarks and Quality Metrics}
We evaluate our work using eleven CUDA kernels from various application domains common to GPUs, in which the occupancy is bounded by the register usage. The first four are graphics kernels. Deferred and SSAO are standard passes used in many modern real time applications. Elevated and Pathtracer are both larger kernels taken from the shadertoys~\cite{shadertoy} web site. Elevated generates an image of a fractal landscape through ray marching using common techniques such as evaluation of fractals and perlin noise. Pathtracer implements a standard path-tracing algorithm.

The other seven kernels are from benchmarks in the Rodinia benchmark suite~\cite{5306797}. They are selected because their occupancy is limited by register pressure, and they are possible to run on the simulator.

Table~\ref{benchmarks} summarizes the kernels, together with their quality metric, their original register usage per thread, and their original occupancy. The graphics kernels all use the Structural Similarity Index (SSIM)~\cite{1284395} to measure quality, which is a well-established metric for comparing the quality of e.g. compressed images. For Hybridsort, we use a binary quality metric, i.e. the output can be correct or wrong. For the remaining kernels, we use percentage of deviation from the correct output. Note that, while SSIM is a well-established quality metric for images, the \% deviation metric might not always be ideal. The choice of quality metric has a large impact on both the possibility to trade bits for output quality, as well as how usable the end result is. Ideally, the quality metric should be decided by application domain experts. However, in this paper, we use it to demonstrate the potential of our approach. The metric can easily be replaced by something more appropriate, without any impact on our approach.

\begin{table}
    \centering
    \caption{A summary of the evaluated kernels.}
    \label{benchmarks}
\resizebox{\linewidth}{!}{
    \begin{tabular}{|c|c|c|c|c|}
    \hline
        & \textbf{Quality}& \textbf{Register usage} & \textbf{Warps} & \textbf{Group}\\
    \textbf{Name}& \textbf{metric} & \textbf{per thread} & \textbf{per block} &\\
    \hline
    Deferred    & SSIM                  & 47 & 8 & 1 \\ \hline
    SSAO        & SSIM                  & 28 & 8 & 1\\ \hline
    Elevated    & SSIM                  & 46 & 8 & 1 \\ \hline
    Pathtracer  & SSIM                  & 50 & 8 & 1\\ \hline
    CFD         & \% deviation          & 60 & 6 & 2\\ \hline
    DWT2D       & \% deviation          & 38 & 6 & 2\\ \hline
    Hotspot     & \% deviation          & 31 & 8 & 2\\ \hline
    Hotspot3D   & \% deviation          & 42 & 8 & 2\\ \hline
    IMGVF       & \% deviation          & 52 & 10& 2\\ \hline
    GICOV       & \% deviation          & 24 & 6 & 2\\ \hline
    Hybridsort  & Binary                & 36 & 8 & 3\\ \hline

    \end{tabular}
    }
\end{table}


\section{Results}
\label{results}
We first investigate what impact the static framework has on the register pressure and occupancy. Second, we examine the performance impact in terms of instructions per clock (IPC). Third, we carry out a sensitivity analysis with respect to writeback delay. Fourth, we present an area overhead analysis.

\subsection{Impact on Register Pressure and Occupancy}
We consider two output quality thresholds. The first one is when \textit{no} quality degradation is allowed, called  \textit{perfect quality}, and define it as SSIM = 1.0 for Group 1 (see Table~\ref{benchmarks}), and as 0\% deviation for Group 2. The metric of Hybridsort is binary and has only two levels: perfect and not acceptable. The second threshold  is when a slight quality loss is accepted. We call this \textit{high quality}, and define it as SSIM = 0.9 for Group 1, 10\% deviation for Group 2, and perfect for Hybridsort (since its quality metric is binary). Up to 10\% quality loss is generally acceptable~\cite{osti_1286958}, but note that this threshold should be carefully considered by the domain expert.

\begin{figure}
\centering
\includegraphics[width=\columnwidth]{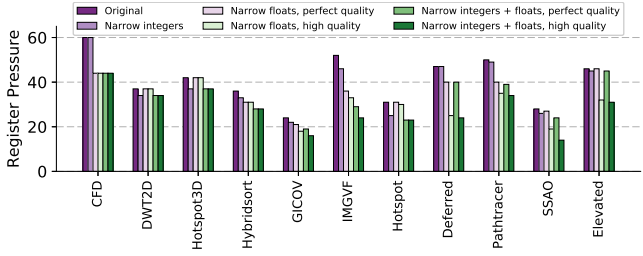}
\caption{Original register pressure and the register pressure when using the static analysis framework for two different quality thresholds.}
\label{regusage}
\end{figure}

Figure~\ref{regusage} presents the impact the static framework has on the register pressure: the y-axis shows the required number of registers per thread, and we present 6 bars per benchmark. The first bar, from the left, shows the original register pressure. The second bar shows the register pressure if only integers are compressed. The third and fourth bars show the register pressure if only floats are considered for compression, for perfect and high quality. The fifth and sixth bar show the register pressure if \textit{both} integers and floats are compressed, for perfect and high quality, respectively. 

The framework reduces the register pressure for all benchmarks. Hybridsort, GICOV, and IMGVF show the largest relative reduction, since they respond well to both parts of the framework. While the floating-point reduction framework is responsible for the largest reduction in register pressure overall, for some benchmarks (e.g. DWT2D, Hotspot3D, Hotspot) the static integer reduction framework is of key importance to achieve a register pressure reduction.

\begin{figure}
\centering
\includegraphics[width=\columnwidth]{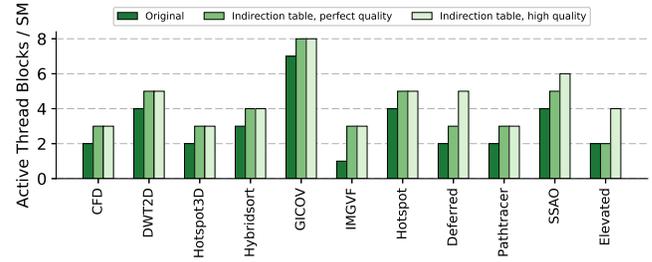}
\caption{Impact on occupancy for two quality thresholds.}
\label{occupancy}
\end{figure}

Figure~\ref{occupancy} presents the impact of the register pressure reduction on the occupancy. The first bar, from the left, shows the original occupancy. The second and third bars show the occupancy when using our proposed approach, for a perfect and a high output quality. Here, the entire framework is used, which means that both integers and floats are reduced. In all cases, the occupancy increases. However in some cases, the decrease in register pressure does not transform into a corresponding increase in occupancy. This is because shared memory-usage can also limit the achievable occupancy. For example, consider the result of IMGVF. When going from perfect to high output quality, the register pressure is reduced from 29 to 24 registers. If only register pressure was the limiting factor, the occupancy would increase to four blocks, since $24$ registers $\times 32$ threads $\times 10$ warps $\times 4$ blocks $= 30 720 < 32 768$. However, each block also uses 14,560 bytes of shared memory, meaning that no more than 3 blocks can fit into the 48 KB shared memory of the SM.

\subsection{Impact on Performance}
\label{IPCimpact}
\begin{figure}
\centering
\includegraphics[width=\columnwidth]{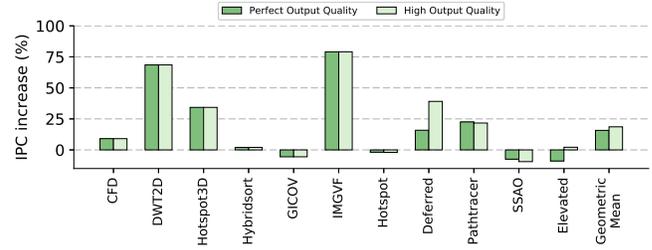}
\caption{Impact on IPC for two quality thresholds.}
\label{performance}
\end{figure}

Figure~\ref{performance} shows the impact on IPC when using the proposed register file organization, for a perfect and high output quality. In many cases, the IPC correlates with the increase in occupancy, with an increase in geometric mean of 15.75\% and 18.6\% for a perfect and high output quality, respectively. For CFD, DWT2D, Hotspot3D, IMGVF, Deferred, and Pathtracer we see a substantial increase in IPC (between 9\% and 79\%). However, for some benchmarks the IPC decreases. For GICOV and SSAO, the IPC decrease is due to contention in the texture cache. For a perfect quality output, the miss rate increases from 76\% to 86\% for GICOV, and from 69\% to 73\% for SSAO, which hurts performance.

For Elevated, there is a decrease in IPC when targeting a perfect output quality, but a slight increase for the high quality output. This is because the new operand collector has a deeper pipeline, which requires more warps. We investigate the relationship between IPC and pipeline stages further in Section~\ref{sensitivity_analysis}.

\subsection{Sensitivity Analysis: Writeback Delay}
\label{sensitivity_analysis}
In the evaluation in Section~\ref{IPCimpact}, we model the writeback delay for \textit{each} operand as three clock cycles; one cycle for conversion from high to low precision, one cycle for accessing the register file, and one cycle to account for possible indirection table bank conflicts. This is quite a pessimistic estimation since not every operand needs to be converted. In addition, the risk for bank conflicts is low since the writeback pipeline is three operands wide, and the number of banks is 16. Nevertheless, the true number of required cycles might be either more or less than three, which motivates a sensitivity analysis of the writeback delay. 

\begin{figure}
\centering
\includegraphics[width=\columnwidth]{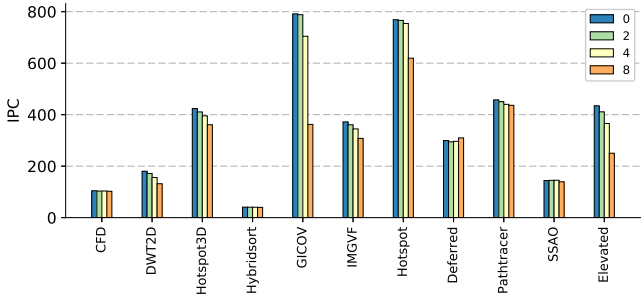}
\caption{Impact on IPC when varying the number of writeback delay cycles.}
\label{writeback_delay}
\end{figure}

Figure~\ref{writeback_delay} shows the resulting IPC for all benchmarks when assuming four different writeback delays: 0, 2, 4, and 8 cycles. For most of the benchmarks, the impact is small up to four cycles. The exceptions are Elevated and GICOV, for which IPC significantly deteriorates at four cycles. The reason is that GPUs do not include forwarding but rely on a scoreboard that prevents scheduling dependent instructions resulting in lower IPC. Note that timing anomalies sometimes give non-intuitive increases of IPC for larger writeback delays, e.g. in Deferred.

\subsection{Area Overhead Analysis}
\label{overhead_analysis}
This section uses transistor count as a proxy for the area overhead of the green blocks in Figure~\ref{newopndcoll}.

The value extractors are the most transistor hungry. Each thread-level value extractor (TVE) consists of eight 9:1-multi-plexers which are 32 bits wide. Assuming each bit of each multiplexer can be implemented with eight 6-transistor AOI cells, the transistor count is 1536 per TVE. Additionally, each TVE also requires one 2:1-multiplexer, four bits wide, which adds $6\times4=24$ transistors. Since each warp consists of 32 threads, each warp-level extractor requires about 50K transistors. In total, this sums up to $50K\times16=800K$ transistors, since one extractor per bank is needed.

We synthesize the Value Converter to the register-transfer level. By analyzing the resulting gate network, consisting mainly of an adder and some multiplexers, we estimate the transistor count per thread-level value converter to be approximately 1300. This sums up to 249,600 transistors for 6 warp-level value converters. Each indirection table has 256 32-bit entries. Assuming each bit is stored with a 6-transistor SRAM cell, the transistor count for each indirection table is 49,152. Two indirection tables are needed: 98,304 transistors in total.

The number of transistors for the value truncators can be estimated using the value converter and value extractor overheads. Each thread-level value truncator consists of one thread-level value truncator and two thread-level value extractors. We assume that a value truncator requires roughly the same area as a value converter, since the steps taken are similar. Then, each thread-level value truncator requires $1 \times 1300 + 2 \times 2048 = 5396$ transistors, in total $5396 \times 32 \times 3 = 518,016$ for three warp-level value truncators.

Finally, the extension for one CU consists of one 1024-bit wide OR-gate and additional SRAM-cells. Assuming a 6-transistor OR-gate per bit, and 35 bits additional storage per CU, the overhead per CU is $1024 \times 6 + 35 \times 3 \times 6 = 6774$ transistors; 108,384 for all 16 units. 

In total, the estimated transistor budget for all structures is about 1.8 million transistors per streaming multiprocessor, i.e around $1,800,000 \times 15 = 27,000,000$ transistors in total. This is a pessimistic estimation, since no circuit-level optimizations have been considered. Still, it is a very small fraction (less than 1\%) of the total transistor budget, which is about 3.1 billion transistors for the GTX 480 chip.

\subsection{Power Overhead Analysis}
\label{power}
We estimate the power overhead analytically by considering static and dynamic power, separately.

Generally, static power increases linearly with the circuit area: hence, we estimate the overhead in static power to increase linearly with the area overhead estimated in the previous section.

When it comes to dynamic power, we compare our proposed design with a twice as big register file. The rationale is that for some benchmarks, our design more than double the number of active thread blocks. To reach the same occupancy by increasing the register file size, it also has to (more than) double. Our conclusion is that our design increases dynamic power less than what a twice as big a register file would do. We come to this conclusion based on three reasons: 

First, the largest difference from the original pipeline is that the proposed design occasionally fetches two registers instead of one during a register read. Naturally, this behaviour increases the dynamic power of the register read by 2x when a double-fetch happens. However, how often this occurs is controlled by the compiler, since it makes the decision whether an operand should i) be split and placed in two different physical registers or ii) be placed contiguously in one physical register. Hence, the compiler could be designed to be aware of the trade-off between minimizing fragmentation (i) and minimizing power dissipation (ii). 

Second, in the worst case, the proposed register file organization increases the number of register fetches by 2x, which would lead to a doubling in power for each register read. However, note that this does not necessarily increase the power more than it would to instead double the size of the register file. Because of the banked register file organization, a doubling in capacity means a doubling in the number of entries of each bank, which means a doubling in bitline length. Since most of the dynamic power consumed in an SRAM is due to bitline charging~\cite{quietbitline}, a doubling in bitline length also doubles the consumed dynamic power per register read.

Finally, as for the rest of the added structures, we estimate their contribution to the dynamic power overhead to be small: the dynamic power from the value converters, value truncators, and value extractors are negligible in comparison to the power consumption of the large register file since the energy per operation is typically an order of magnitude below that of SRAM structures~\cite{7479518}. Furthermore, while the indirection tables are SRAM structures, they are also very small in comparison to the register file.

\section{Discussion}
\label{discussion}

Our evaluation uses the NVIDIA Fermi architecture as baseline. However, it is important to understand the design implications on newer architectures. Here, we give a comparison on how it scales to the NVIDIA Volta~\cite{voltawhitepaper} architecture.

Register shortage is a reality also in newer architectures. While the total register-file size of Volta is much larger than Fermi (20480 KB vs 1965 KB), the Volta architecture also supports more threads in total: each thread only has 31 32-bit registers available at maximum occupancy. Keeping the register count low continues to be a problem for programmers, as the tuning has to be carried out manually~\cite{nvidia_best_practices_guide}. This is a cumbersome task which requires the programmer to either re-write the kernel or accept inefficient trade-offs such as register spilling to reach a high occupancy. Hence, register shortage remains a problem which can be alleviated by employing our approach.

When it comes to area, our estimate is that the overhead is slightly larger for Volta than for Fermi, but still very small (just over 2\% of the total transistor budget). Note that this is a pessimistic estimate, since no optimizations has been considered, as further discussed below. The main reason for the increase is that Volta has a higher count of individual register files than Fermi. We derive this conclusion from the discussion below.

Recall from Section~\ref{gpu_background} that Fermi has two warp schedulers per SM. They share a register file of 256 KB, as well as all the computing units in the SM. In comparison, each SM in the Volta architecture is partitioned into four processing blocks, where each block has a dedicated warp scheduler, a register file of 64 KB, and its own computing units. Each register file requires its own operand collector, and thus dedicated indirection tables, value extractors, and value converters. 

Since the number of register banks scales with the maximum instruction throughput (two per SM for Fermi vs one per processing block for Volta), and we need one value extractor per bank, we assume that Volta requires half of the value extractors needed for a Fermi register file, which corresponds to 400k transistors according to Section~\ref{overhead_analysis} . Assuming all other structures are unchanged, we get an area overhead of 1.8M-0.4M = 1.4M transistors per processing block, or 5.6M transistors per SM. The Volta architecture has 84 SMs, which sums up to a total area overhead of ~470 million transistors. Although this is a higher count than for the Fermi architecture, Volta also has a significantly higher transistor count of 21 billion transistors in total. As a result, the area overhead is still a very small fraction compared to the total transistor budget (just over 2\%). Note that this is a pessimistic estimation, since no circuit-level optimizations has been considered. In addition, while out of scope for this paper, it might be possible to share some of the structures between the processing blocks, which would further reduce the area overhead.

\section{Related Work}
\label{relatedwork}
GPU register file optimizations have been addressed in several prior studies. Gilani et al.~\cite{6522330}, Esfeden et al.~\cite{AsghariEsfeden:2019:CCO:3297858.3304026}, as well as Wang and Zhang~\cite{8029511} investigate optimizations based on narrow integers which are detected at run time, in stark contrast to our static approach which works for all types of narrow data. Voitsechov et al.~\cite{Voitsechov:2018:STI:3274266.3243905} employ narrow integer-packing based on static analysis, but they do not support floats.

Angerd et al.~\cite{Angerd:2017:FAC:3154814.3151032} present a study on reducing the bitwidth of floating-point values, which is the method we adopt to tune the precision of floats offline. Their method assumes an indirection table capable of handling floating-point operands of different bitwidths, but they do not present a microarchitecture design of the register file nor a complete register-file design capable of dealing with both integer and floating-point operands.  We present a complete design, at the microarchitecture level, of such a register file.

Other related studies include Jeon et al.~\cite{Jeon:2015:GRF:2830772.2830784} who investigate releasing dead registers and re-allocating them to other warps. Yu et al.~\cite{YU201625} propose a technique which increases the number of active warps by employing run-time allocation. Furthermore, Khorazani et al.~\cite{regmutex} propose a software-hardware co-mechanism where some operands are statically allocated, while others time-share registers. Also, RegLess~\cite{Kloosterman:2017:RJO:3123939.3123974} uses a compiler-supported technique to only allocate register file space to currently accessed regions of code, a technique orthogonal to our approach. All these techniques are orthogonal to our, since they do not target reduction of register pressure.

Furthermore, Lee et al.~\cite{Lee:2015:WEP:2749469.2750417} target register compression to enable more power-efficient GPUs. While this technique might lower the physical register usage per thread, their microarchitectural implementation specifically targets power consumption, while we target performance improvements.

\section{Conclusion}
\label{conclusion}
Modern GPUs rely on TLP to provide high throughput. The thread register footprint limits TLP, since the state of all active threads must be readily available in the register file. In this paper, we propose a new concept for efficient register-packing, which combines static integer and float operand compression with a novel GPU register file organization capable of lowering the register footprint by densely storing narrow operand values. We present a detailed microarchitectural implementation of the proposed organization, together with a performance evaluation and an overhead analysis. Our results show that the IPC of the investigated benchmarks can be increased by up to 79\%, 18.6\% on average, when allowing for a slight quality output degradation. 

\section*{Acknowledgments}
This work is supported by the Swedish Research Council under contract numbers VR-2014-06221 and VR-2019-04929.
\bibliographystyle{ACM-Reference-Format}
\bibliography{ref}

\end{document}